\documentclass[journal]{IEEEtran}

\makeatletter
\def\endthebibliography{%
	\def\@noitemerr{\@latex@warning{Empty `thebibliography' environment}}%
	\endlist
}
\makeatother

\usepackage{cite}

\usepackage[pdftex]{graphicx} 

\usepackage[caption=false,font=footnotesize]{subfig}
\usepackage[export]{adjustbox}

\usepackage{amsmath}
\usepackage{amsfonts}
\usepackage{bm}

\usepackage{filecontents}
\usepackage{amssymb}
\usepackage{color}

\usepackage{epsfig}

\usepackage{verbatim}

\usepackage{url}

\usepackage{hyperref}

\usepackage{algorithm, tabularx}

\usepackage{lettrine}

\usepackage{lipsum}

\usepackage{siunitx}

\usepackage{soul,color}

\usepackage{mathrsfs}

\usepackage{array}

\usepackage[inline]{enumitem}

\usepackage{dblfloatfix} 

\usepackage{algorithm}

\usepackage[noend]{algpseudocode}

\makeatletter
\DeclareRobustCommand{\iscircle}{\mathord{\mathpalette\is@circle\relax}}
\newcommand\is@circle[2]{%
  \begingroup
  \sbox\z@{\raisebox{\depth}{$\m@th#1\bigcirc$}}%
  \sbox\tw@{$#1\square$}%
  \resizebox{!}{\ht\tw@}{\usebox{\z@}}%
  \endgroup
}
\makeatother

\usepackage{setspace}
\usepackage{multirow}
\usepackage{array}
\newcolumntype{L}[1]{>{\raggedright\let\newline\\\arraybackslash\hspace{0pt}}m{#1}}
\newcolumntype{C}[1]{>{\centering\let\newline\\\arraybackslash\hspace{0pt}}m{#1}}
\newcolumntype{R}[1]{>{\raggedleft\let\newline\\\arraybackslash\hspace{0pt}}m{#1}}

\newcommand{\ie}{\textit{i}.\textit{e}.}

\begin{document}
	
\title{AI-Enhanced Wide-Area Data Imaging via Massive Non-Orthogonal Direct Device-to-HAPS Transmission}




	
	






	\author{Hyung-Joo Moon,~\IEEEmembership{Graduate Student Member,~IEEE}, Chan-Byoung Chae,~\IEEEmembership{Fellow,~IEEE},\\ Kai-Kit Wong,~\IEEEmembership{Fellow,~IEEE}, and  Robert W. Heath, Jr.,~\IEEEmembership{Fellow,~IEEE}
        \thanks{H.-J. Moon and C.-B. Chae are with the School of Integrated Technology, Yonsei University, Seoul 03722, South Korea (e-mail: \{moonhj, cbchae\}@yonsei.ac.kr).}
		\thanks{K.-K. Wong is with the Department of Electronic and Electrical Engineering, University College London, London WC1E 7JE, UK. He is also affiliated with Yonsei Frontier Lab., Yonsei University, Seoul 03722, South Korea (e-mail: kai-kit.wong@ucl.ac.uk).}
		\thanks{R. W. Heath, Jr. is with the Department of Electrical and Computer Engineering, University of California at San Diego, La Jolla, CA 92093, USA (e-mail: rwheathjr@ucsd.edu).}
	}
	
	

	\maketitle

	\begin{abstract}

	Massive Aerial Processing for $X$ (MAP-$X$) is an innovative framework for reconstructing spatially correlated ground data, such as environmental or industrial measurements distributed across a wide area, into data maps using a single high altitude pseudo-satellite (HAPS) and a large number of distributed sensors. With subframe-level data reconstruction, {MAP-$X$} provides a transformative solution for latency-sensitive IoT applications. This article explores two distinct approaches for AI integration in the post-processing stage of MAP-$X$. The DNN-based pointwise estimation approach enables real-time, adaptive reconstruction through \emph{online} training, while the CNN-based image reconstruction approach improves reconstruction accuracy through \emph{offline} training with non-real-time data. Simulation results show that both approaches significantly outperform the conventional inverse discrete Fourier transform (IDFT)-based linear post-processing method. Furthermore, to enable AI-enhanced MAP-$X$, we propose a ground-HAPS cooperation framework, where terrestrial stations collect, process, and relay training data to the HAPS.
    With its enhanced capability in reconstructing field data, AI-enhanced MAP-$X$ is applicable to various real-world use cases, including disaster response and network management.


	\end{abstract}

	\IEEEpeerreviewmaketitle

	\section{Introduction}


\IEEEPARstart{T}{he} exponential growth of connected devices has established the Internet of Things (IoT) as a foundational technology in modern digital infrastructure~\cite{oacref}. IoT networks now play a critical role in wide-area monitoring by leveraging spatiotemporal correlations in industrial and environmental data to facilitate real-time intelligence and decision-making. Existing IoT solutions, however, still face challenges in coverage, data acquisition speed, and connectivity, particularly in vast or remote areas. Addressing these limitations requires breakthrough wide-area sensing technologies that provide broader coverage, ultra-low latency, and scalable support for massive device access.



\begin{figure}[t]
	\begin{center}
		{\includegraphics[width=0.9\columnwidth,keepaspectratio,frame]
			{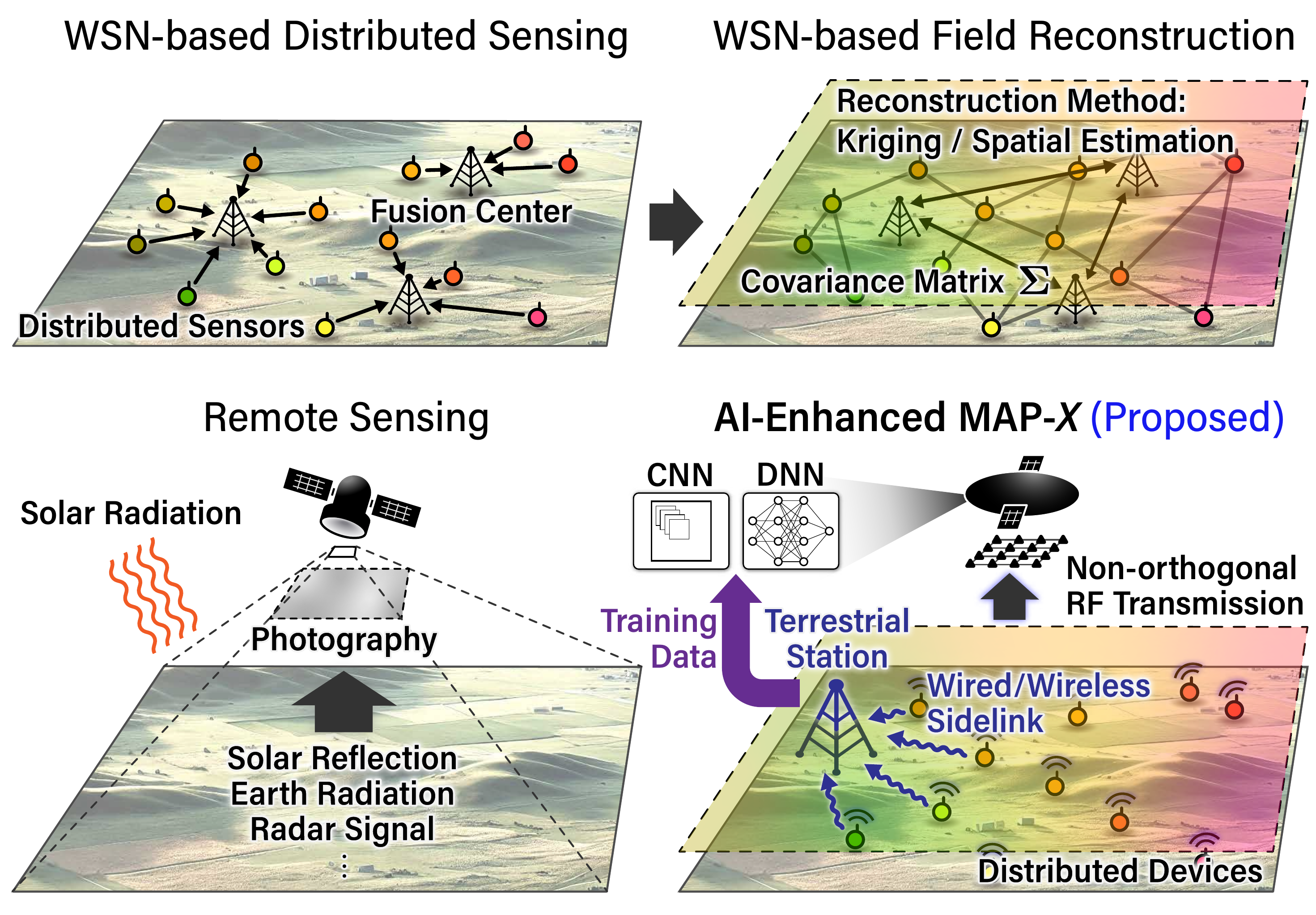}%
			\caption{Comparison of traditional GIS (WSN-based collection and remote sensing) and MAP-$X$ with deep learning integration. The potential use cases of MAP-$X$ are illustrated in Fig.~\ref{fig04}.}
			\label{fig01}
		}
	\end{center}
	\vspace{-15pt}
\end{figure}

A geographic information system (GIS), which maps geographically referenced information, is a key objective of wide-area sensing technologies~\cite{gisref}. Traditionally, GIS implementations have relied on wireless sensor networks (WSNs) and remote sensing. The WSN-based approach reconstructs extensive field data using techniques such as kriging interpolation or spatial estimation from distributed sensor measurements, while remote sensing employs satellite or aerial imagery to detect solar reflection or terrestrial radiation. Fig.~\ref{fig01} illustrates three wide-area sensing frameworks: the WSN-based method, remote sensing, and the proposed artificial intelligence (AI)-enhanced MAP-$X$ framework.



Each conventional approach offers distinct advantages and limitations. Satellite remote sensing enables large-scale environmental monitoring but often depends on indirect inference, as it cannot directly measure certain parameters~\cite{remsenref}. Other strategies, such as UAV-based sensing and hybrid ground-air architectures, provide mobility and flexibility but face limitations in coverage, endurance, or latency due to flight constraints and multi-hop coordination. In contrast, WSNs facilitate targeted data collection through distributed sensors that relay specific measurements to a fusion center. However, the limited communication range of low-power sensors makes direct transmission to the fusion center impractical. Consequently, large-scale deployments rely on multi-hop networking, which introduces high latency and complex scheduling challenges. Such limitations are particularly critical for applications such as disaster response and network management, where wide-area sensing, low latency, and extensive connectivity are essential.

High altitude pseudo-satellites (HAPS) equipped with multiple-input multiple-output (MIMO) or reconfigurable intelligent surfaces (RIS) are expected to play a key role in future non-terrestrial networks (NTNs), providing active spatial processing at altitude~\cite{myjref,risref}. MAP-$X$ leverages a non-orthogonal, direct device-to-HAPS transmission scheme for wide-area and low-latency GIS. In this framework, numerous ground devices simultaneously transmit unique multi-carrier radio-frequency (RF) waveforms over shared time-frequency resources. These superimposed signals are captured by a HAPS equipped with a uniform planar array (UPA) antenna. Using angle-of-arrival (AoA) domain processing, MAP-$X$ superimposes signals from nearby devices while spatially distinguishing those from more distant sources. Given that environmental and industrial data typically exhibit high spatial correlation, adjacent devices transmit nearly identical data, leading to overlapping signals that reinforce each other. This increases the average signal-to-noise ratio (SNR), making direct transmission from low-power devices to the HAPS more efficient. Mathematical analysis in~\cite{myjref} confirms that effective data reconstruction is attainable using a simple inverse discrete Fourier transform (IDFT)-based linear post-processing method.


The performance of MAP-$X$ can be further enhanced through deep-learning integration. 
Building on the IDFT-based linear post-processing model, deep neural networks (DNNs) can be used for adaptive filtering and soft clipping. These DNNs are trained online using local device data aggregated via terrestrial stations, ensuring real-time adaptation to fluctuating environmental conditions. Additionally, a convolutional neural network (CNN)-based approach is proposed, which processes image-type inputs and outputs using batch-collected training data for offline training. Simulation results demonstrate that these AI-based methods significantly outperform the HAPS-standalone IDFT-based approach, highlighting the value of AI-driven processing and the importance of collecting and relaying training datasets to the HAPS.

	\section{MAP-$X$ for Real-Time Wide-Area Data Imaging}

In this section, we introduce the MAP-$X$ framework by detailing the five main components and the operational phases that enable real-time wide-area data imaging. We begin by outlining the system model, including the sensing target, devices, non-terrestrial station, terrestrial station, and environments. We then describe the three sequential communication phases of MAP-$X$. Finally, we discuss the key innovations in waveform design and transmission strategy that empower the system.

	\subsection{System Model of MAP-$X$}

	\subsubsection{Sensing Target}


Many environmental and industrial monitoring tasks aim to measure physical phenomena, referred to here as the sensing targets. These targets often exhibit both temporal and spatial correlations, allowing traditional methods such as remote sensing and WSNs to reconstruct data maps~\cite{sblueref2}. However, remote sensing is limited by the types of data it can directly observe, while WSNs are constrained by communication capacity and power. As a result, both approaches struggle with real-time reconstruction of rapidly evolving field data. MAP-$X$ addresses this challenge by enabling precise data reconstruction within milliseconds, allowing accurate and immediate interpretation of dynamic wide-area phenomena. Consequently, the sensing targets of MAP-$X$ include not only slowly varying environmental conditions, but also diverse time-sensitive metrics.
	
	\subsubsection{Devices}
	
Similar to traditional distributed sensing frameworks based on WSNs, MAP-$X$ utilizes devices that both sense target data and transmit low-power signals~\cite{sblueref2}. These devices remain in an idle communication state until they receive a trigger signal from the HAPS. They are required to have RF-band transceivers with a minimum bandwidth of several tens of kHz and a high-precision local clock. Accordingly, MAP-$X$ can utilize a range of devices, from dedicated sensors with communication capabilities to smartphones endowed with sensing functionalities.

\begin{figure*}[t]
	\begin{center}
		{\includegraphics[width=1.9\columnwidth,keepaspectratio,frame]
			{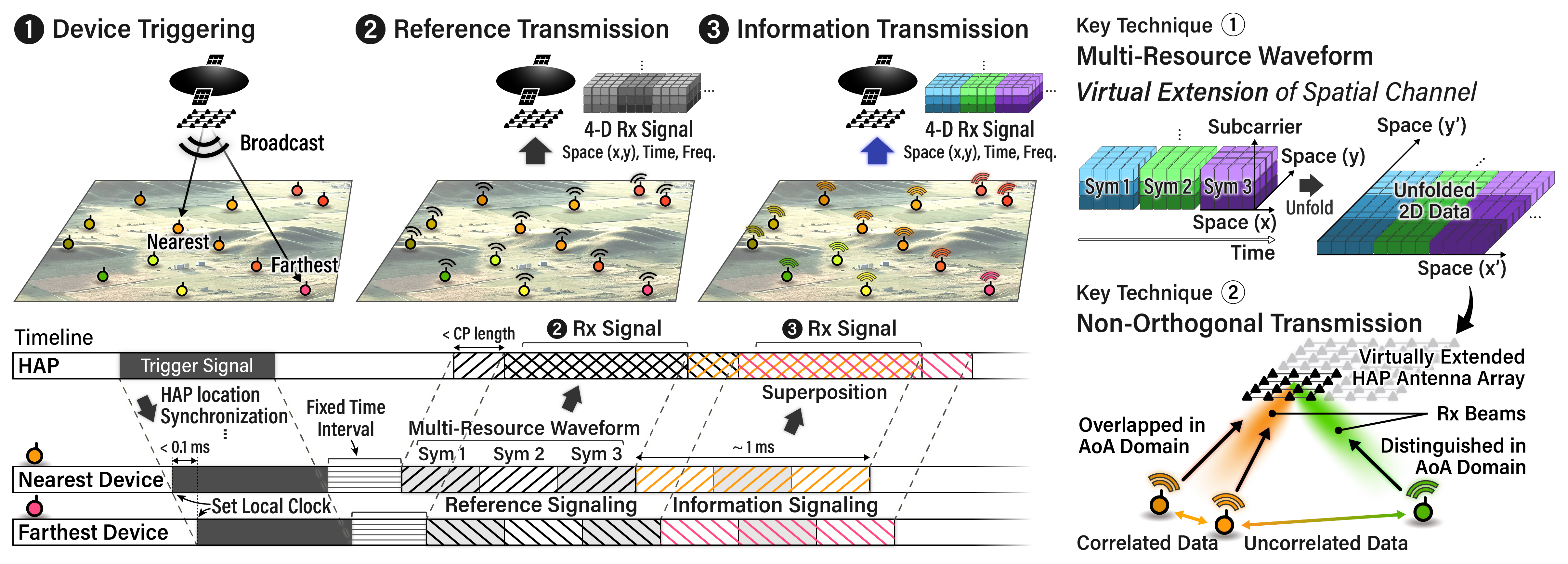}%
			\caption{Three transmission phases of MAP-$X$ and the two key techniques supporting its operation.}
			\label{fig02}
		}
	\end{center}
	\vspace{-15pt}
\end{figure*}

	\subsubsection{Non-Terrestrial Station}
	
The non-terrestrial station in MAP-$X$ can be implemented using satellites or various types of high-altitude platforms. In this work, we adopt an aerostatic HAPS as the primary platform. These platforms offer quasi-stationary positioning, suitable coverage range for MAP-$X$, minimal Doppler shifts, and sufficient payload capacity to support the large UPA and the respectively connected RF chains. The HAPS fusion center captures RF signals transmitted from the devices through the UPA. Although the non-orthogonal transmission scheme prevents the receiver from isolating individual device signals, the high altitude of the HAPS enables mapping the superimposed signals from specific locations onto the AoA domain channel. By leveraging a uniquely designed waveform and an appropriate post-processing strategy, target data can be reconstructed within the two-dimensional (2D) ground spatial domain.

	\subsubsection{Terrestrial Station}
	
While the terrestrial station is not involved in the real-time transmission and reception process of MAP-$X$, it is critical for supporting the AI-enhanced capabilities. Depending on the model, the training data consist of either location and measurement pairs for each sensor or precisely reconstructed full data images generated without real-time constraints. Therefore, the terrestrial station is responsible for gathering and processing these datasets for training.

	\subsubsection{Environments}
	
MAP-$X$ is designed to operate reliably even in challenging environments with non-uniform device distribution and complex topography~\cite{myjref}. The system experiences only minor performance degradation in non-line-of-sight (NLoS) conditions, primarily due to SNR reduction. The system requires channel coherence on the order of several milliseconds, which is a realistic assumption given the low mobility of sensing devices and the pseudo-stationary position of the HAPS. Moreover, because MAP-$X$ does not require detailed channel estimation for each individual device-to-HAPS link, it avoids performance issues related to channel estimation errors or feedback overhead.


	\subsection{Three Transmission Phases of MAP-$X$}
	
	
MAP-$X$ can be viewed as an efficient data collection and reconstruction strategy within a WSN that employs direct and non-orthogonal transmission. Fig.~\ref{fig02} illustrates the three communication phases, which are described below.
	
	\subsubsection{Device Triggering}
	
The process begins when the HAPS broadcasts a trigger signal to activate the devices. This signal conveys essential information, including the type of data to be collected, the HAPS's location, and the distance between the HAPS and the nearest device. Additionally, the trigger signal synchronizes the devices’ local oscillators and clocks, ensuring time alignment based on their distance from the HAPS. For instance, the nearest device maintains the earliest clock timing, while the farthest device operates on the latest clock timing.
	
	\subsubsection{Reference Transmission}
	
Each device generates a waveform based on its sensing data, location, and the information received from the HAPS, adhering to a specified protocol. Using a uniquely designed waveform based on orthogonal frequency division multiplexing (OFDM) waveform, devices modulate the phases of symbols across multiple time-frequency resources (\ie, subframe) using a predefined rule. The amplitudes of all symbols are at a constant reference value. The rationale for utilizing multiple resource elements in both reference and information transmission phases is detailed in Section~\ref{sec2c}.
	
	\subsubsection{Information Transmission}
	
After completing the reference transmission, devices move to the subsequent subframe designated for information transmission. The phase modulation of symbols follows the same rule as in reference transmission. However, the common amplitude across all symbols now represents the sensing data. Each device encodes its raw data into a positive real valued amplitude using a predefined encoding function. Because neighboring devices tend to observe similar physical phenomena, their encoded amplitudes will be nearly identical during this phase. Assuming that both transmission phases occur within the channel coherence time, the reference transmission provides an estimate of the aggregated channel response from devices located within a specific AoA region. Then, by dividing the received signal in the information transmission phase by that of the reference transmission, the receiver can recover the encoded sensing data for that region through AoA-domain signal processing.

	\subsection{Key Techniques Underlying the Transmission Phases}
    \label{sec2c}
	
The main innovation of MAP-$X$ lies in its novel waveform design and non-orthogonal transmission strategy. The mathematical foundations and performance analysis of these techniques are detailed in~\cite{myjref}. This section focuses on the technical implications and practical effects of these advancements.

	\subsubsection{OFDM-based Multi-Resource Waveform}
	

One of the primary challenges in aerial platforms is the physical limitation on the size of the antenna array, imposed by weight and energy constraints. To address this, MAP-$X$ employs an innovative waveform that uses additional time-frequency resource elements to effectively “extend” the spatial domain resolution. In both reference and information transmission steps, devices modulate OFDM symbol phases, which are computed using a predefined formula based on their precise location relative to the HAPS.
This allows the received symbols in the time-frequency domain to be unfolded into a virtually extended spatial domain, as illustrated in Fig.~\ref{fig02}. As a result, this technique simulates the effect of an extended antenna array, significantly enhancing the resolution of AoA domain channelization. Narrowing the AoA domain channel increases the uniformity of data from a single AoA region, thereby maximizing the effectiveness of the superposition-based transmission scheme.

\begin{figure*}[t]
	\begin{center}
		{\includegraphics[width=1.9\columnwidth,keepaspectratio,frame]
			{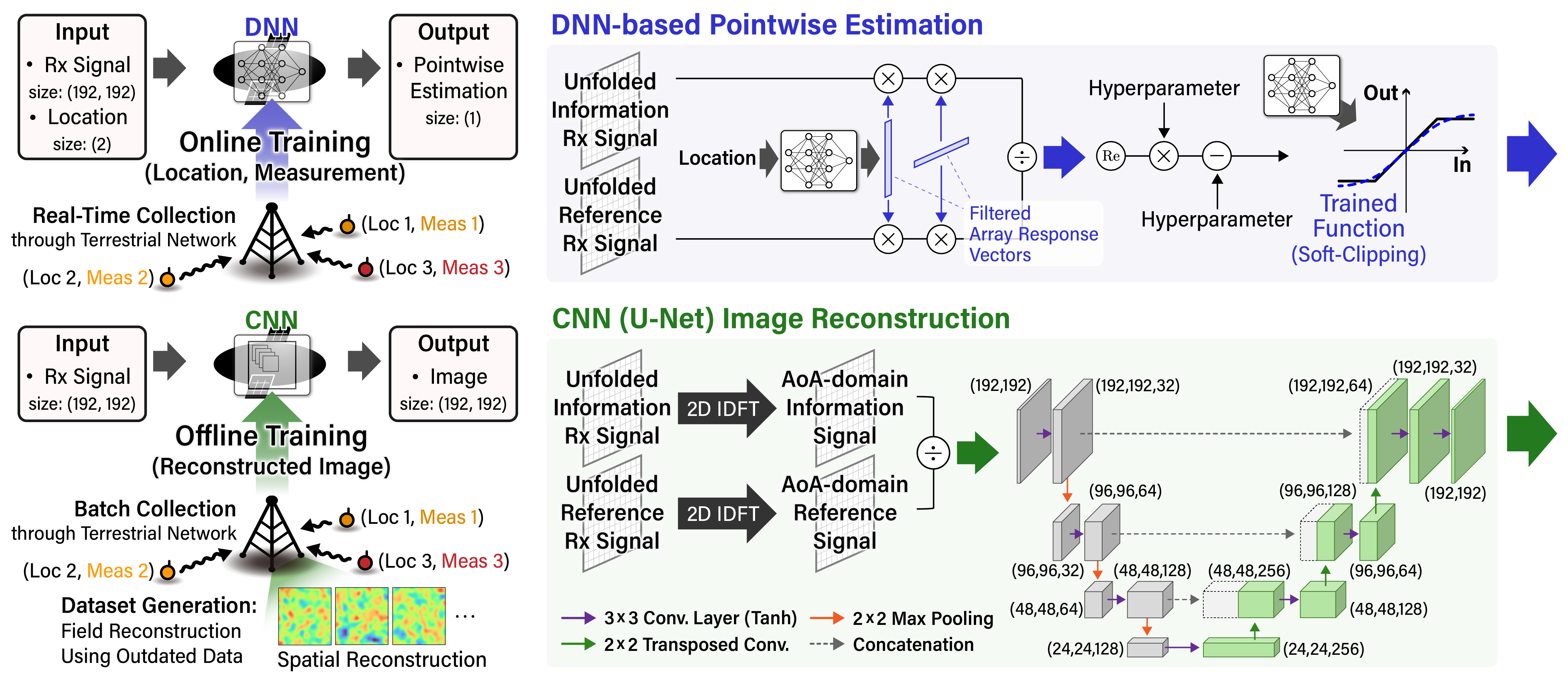}%
			\caption{Training and inference processes of the two deep learning-based post-processing methods.}
			\label{fig03}
		}
	\end{center}
	\vspace{-15pt}
\end{figure*}

	\subsubsection{Overlapping Signals via Non-Orthogonal Transmission}

As outlined in the three transmission phases of MAP-$X$, all devices transmit signals over the same multiple time-frequency resources, leading to signal superposition at the receiver. A related technique, over-the-air computation (OAC), intentionally exploits signal overlap from different devices to perform nomographic operations~\cite{oacref}. MAP-$X$ also leverages signal superposition to ensure that information from closely located devices is effectively averaged, while signals from spatially distant devices remain fully distinguishable in the AoA domain, as illustrated in Fig.~\ref{fig02}. Unlike OAC, which requires uplink channel state information at the transmitters to achieve phase and magnitude alignment at the receiver, MAP-$X$ avoids this requirement entirely, as it does not rely on such alignment. This non-orthogonal transmission strategy offers several key advantages:


$\bullet$ Overlapping data from adjacent devices exploits spatial correlation in the raw data, improving accuracy.

$\bullet$ Non-orthogonal transmission significantly shortens transmission time compared to data gathering through orthogonal data transmission.

$\bullet$ The SNR enhancement from signal overlap enables direct device-to-HAPS transmission, mitigating the high pathloss.

$\bullet$ Under the channel coherence time assumption, MAP-$X$ operates without requiring channel knowledge or individual data decoding, making it resilient to incoherent signal superposition.

\section{AI-Enhanced Post-Processing for MAP-$X$}

Upon completing the three transmission phases of MAP-$X$, the HAPS fusion center acquires two received symbol tensors: one from the reference transmission and another from the information transmission (see Fig.~\ref{fig02}). These tensors are four-dimensional, spanning the spatial $x$- and $y$-axes (antenna array coordinates), time (symbols), and frequency (subcarriers). As described in the previous section, with our waveform design, data in the time and frequency domains can be unfolded into a 2D virtually extended spatial domain. Based on the mathematical derivations in~\cite{myjref}, field reconstruction can be performed by transforming the unfolded data into the AoA domain using a 2D IDFT, followed by division and clipping. This linear reconstruction method is a HAPS-standalone approach, relying solely on the received signals. Furthermore, by utilizing training datasets collected via terrestrial networks, this basic model can be extended to online and offline deep learning-based models for enhanced performance.

\subsection{DNN-based Pointwise Estimation Method}

The DNN-based pointwise estimation method extends a mathematically proven linear estimator by incorporating a trained filter and a trained soft-clipping function to adapt to signal characteristics. The output is designed to estimate target data at a specific location, rather than generating a full data map. Consequently, the input consists of both the received signal and the location to be estimated. As illustrated in Fig.~\ref{fig03}, during signal transformation into the AoA domain, the model is trained to apply adaptive filtering. Additionally, soft-clipping replaces the hard-clipping used in the linear method, resulting in improved estimation accuracy. Because both the filter design and soft-clipping tasks are relatively simple, they can be implemented using small multi-layer perceptron (MLP) models. In our simulations, the filter design network consists of four layers, including the input and output layers, with dimensions 2, 96, 192, and 192, and uses ReLU activation. The soft-clipping network also consists of four layers with dimensions 1, 3, 3, and 1, using a Tanh activation function.

Considering the smaller model size and the availability of real-time training data, online training is feasible for the DNN-based pointwise estimation method. While the HAPS can independently acquire the received signals, the location and raw measurement pairs required for training must be obtained through terrestrial networks. As shown in Fig.~\ref{fig03}, terrestrial stations in the MAP-$X$ framework collect training datasets via sidelink communication between devices and terrestrial stations, rapidly relaying them to the HAPS for online training. This approach enables real-time adaptation to signal characteristics, with the terrestrial station functioning as a transparent relay node, conveying collected data to the HAPS without additional processing.

\subsection{CNN-based Image Reconstruction Method}

The CNN-based image reconstruction method differs from the pointwise estimation method by reconstructing a data map for the entire area, rather than providing location-specific estimations. Building on the IDFT-based reconstruction method, this approach incorporates a U-Net structure for image-to-image processing~\cite{unetref}. As illustrated in Fig.~\ref{fig03}, the process begins with data reconstruction in the AoA domain via a 2D IDFT and division. The U-Net CNN model then performs a nonlinear coordinate transformation from the AoA domain to the 2D ground spatial domain, while also adapting to signal characteristics and denoising the image. The U-Net model architecture is detailed in Fig.~\ref{fig03}.

The input in this case consists only of the received signals, while the output is a fully reconstructed map. Thus, the image reconstruction model must have access to a ground-truth data map that corresponds to the received signals. The terrestrial station plays a crucial role by gathering local data from as many devices as possible via sidelink communication and performing field reconstruction using the collected data without real-time constraints. This dataset generation process is equivalent to a conventional terrestrial WSN-based GIS. Throughout the simulation, we assume that field reconstruction is conducted using the optimal covariance-based spatial estimator~[6]. The resulting dataset can then be used to train the CNN model, either on the HAPS or at the terrestrial station, with the trained model subsequently deployed to the HAPS for inference. Unlike the DNN-based pointwise estimation method, the larger model size and the difficulty of acquiring real-time ground-truth map data restrict the CNN-based image reconstruction method to offline training.


\begin{figure}[t]
	\begin{center}
		{\includegraphics[width=0.9\columnwidth,keepaspectratio,frame]
			{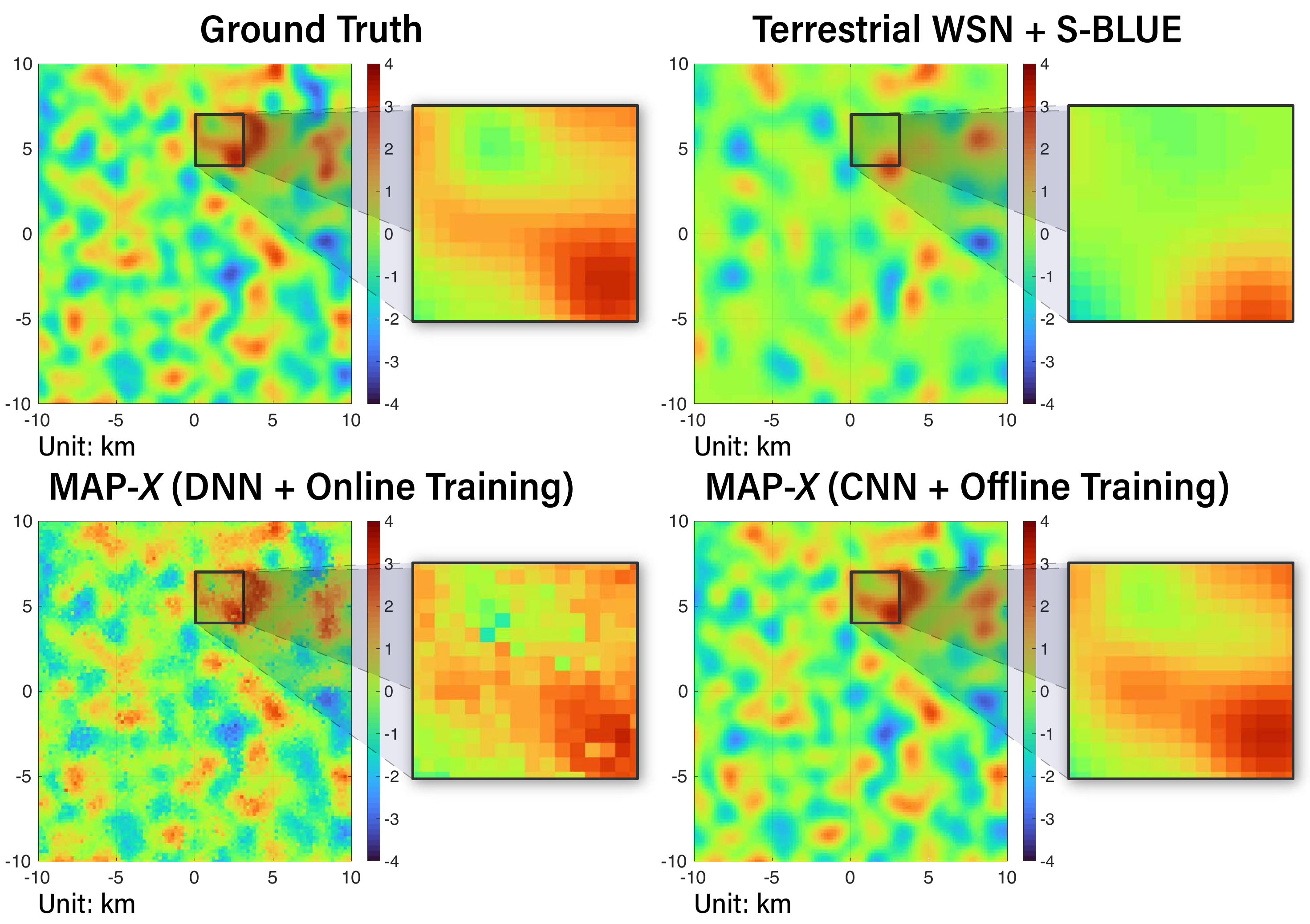}%
			\caption{Image reconstruction results of the conventional WSN-based method and deep learning-aided MAP-$X$ with equivalent resource consumption.}
			\label{fig06}
		}
	\end{center}
	\vspace{-15pt}
\end{figure}

	\subsection{Capabilities of AI-Enhanced MAP-$X$}

We compare MAP-$X$, integrated with linear, DNN-based pointwise estimation, and CNN-based image reconstruction post-processing strategies, against the conventional WSN-based data reconstruction method. In the WSN-based approach, a ground fusion center assigns each device a unique time-frequency resource element to transmit its location and measurement as a single symbol, without requiring prior signaling. The collected data are then used to reconstruct the continuous 2D field using the S-BLUE method, an optimal linear estimator that assumes perfect knowledge of the first- and second-order statistics of the spatial signal~\cite{sblueref2}. Based on this knowledge, S-BLUE computes optimal coefficients to fuse the individual observations into a field estimate. As a result, the WSN-based method represents a highly idealized baseline. In contrast, MAP-$X$ requires no prior statistical knowledge of the data, making it significantly more practical and scalable for real-world deployments.

In the simulation, we consider a total of 50,000 devices distributed over a 40~km~$\times$~40~km area, resulting in a device density of 31.25 devices per km$^2$. 
The raw target data in the simulation are modeled as a 2D Gaussian random field with normalized variance, filtered by a Gaussian function~\cite{myjref}. This filtering yields a correlation coefficient between two data points that decays proportionally to a Gaussian function of their separation distance. The HAPS is equipped with a 16~$\times$~16 antenna array, and each subframe consists of 12 symbols and 12 subcarriers. Each device has a single isotropic antenna; the transmit signal is centered at 2.5~GHz, with a transmission power of 0~dBm. The direct device-to-HAPS transmission channel is modeled as Rician fading.

\subsubsection{Image Reconstruction Results}

Fig.~\ref{fig06} presents the field reconstruction results produced by MAP-$X$ using deep learning-based post-processing methods. In our simulations, four pairs of subframes are used to generate four independent estimates of the ground-truth data map, which are then averaged to produce a single, high-accuracy result. For a fair comparison, eight subframes are allocated to the WSN-based data collection method. For the DNN-based pointwise estimation method, values across the entire field are obtained by repeating the estimation process over a grid of spatial coordinates. This method achieves significantly higher accuracy than the WSN-based approach. However, due to the division step in the estimation process, it exhibits a heavy-tailed error distribution that results in salt-and-pepper noise. This noise can be effectively mitigated by the CNN model, which not only improves estimation accuracy but also recovers finer signal details. Overall, these results demonstrate that MAP-$X$ can accurately reconstruct field data, making it suitable for a wide range of applications, as further discussed in Section~\ref{sec4}.


\subsubsection{Latency vs. Accuracy Performance Analysis}

\begin{figure}[t]
	\begin{center}
		{\includegraphics[width=0.9\columnwidth,keepaspectratio,frame]
			{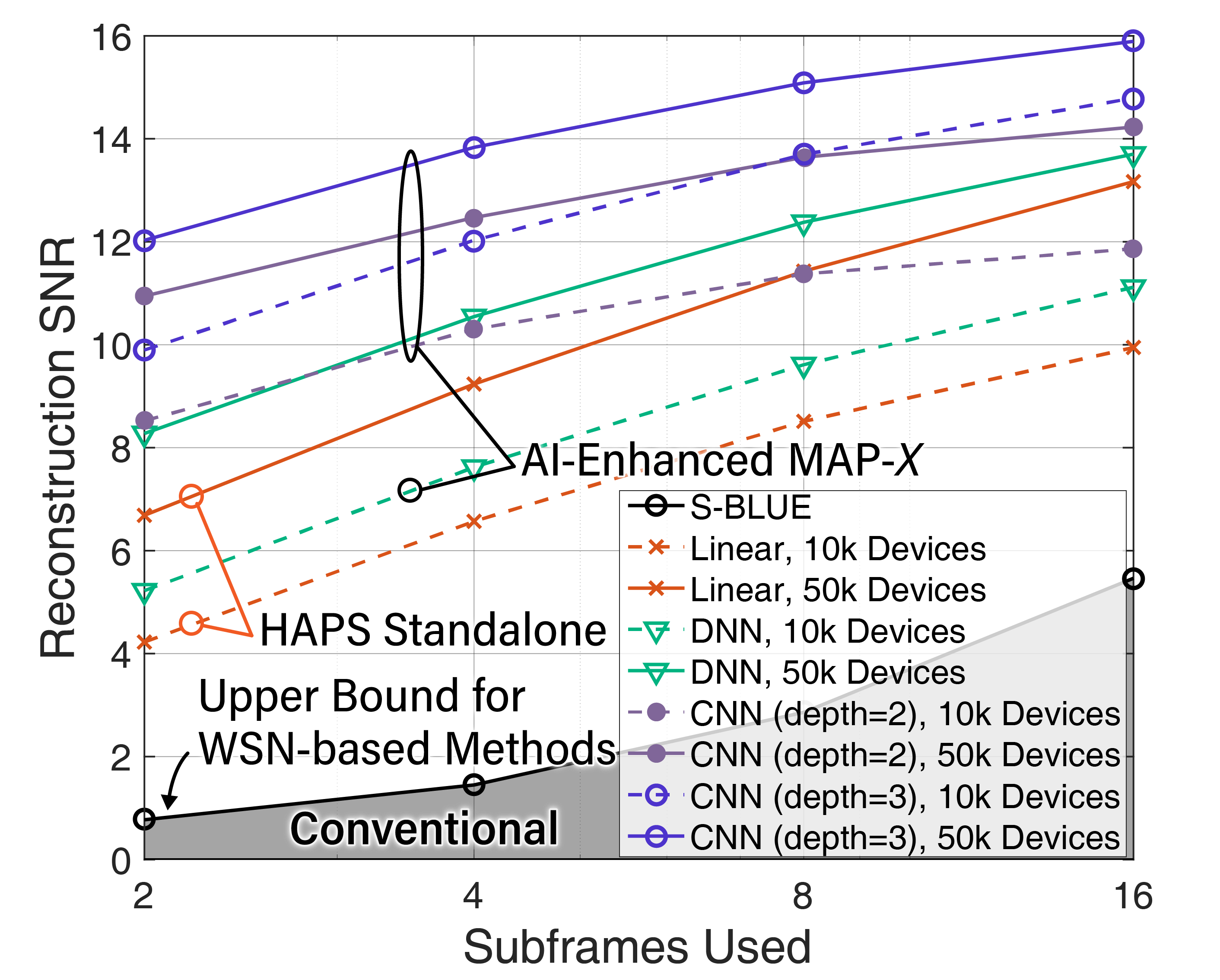}%
			\caption{Comparison of reconstruction SNR performance between the conventional WSN-based method, the HAPS standalone method, and the methods incorporating deep learning models. The black solid line represents the upper bound performance of conventional distributed sensing with orthogonal transmission.}
			\label{fig05}
		}
	\end{center}
	\vspace{-15pt}
\end{figure}

Fig.~\ref{fig05} compares the performance of MAP-$X$ with conventional WSN-based field reconstruction. The $x$-axis represents the number of subframes used to reconstruct a single data map. S-BLUE utilizes additional subframes to gather more device data, while MAP-$X$ employs a scalable approach, averaging the results of multiple MAP-$X$ processes. As shown in Fig.~\ref{fig05}, the upper bound performance of the traditional WSN-based method is significantly lower than the performance achieved by MAP-$X$. The DNN trained through online training improves performance, while the CNN, which requires offline training, achieves much higher accuracy under the same latency conditions. The pointwise estimation method is preferable for scenarios with rapidly changing signal characteristics and limited training data, whereas the image reconstruction method is more effective when target data exhibit strong temporal correlation and ample datasets are available from the terrestrial station.

	
	

\section{Potential Use Cases of MAP-$X$ and Future Research Directions}
\label{sec4}

\begin{figure*}[t]
	\begin{center}
		{\includegraphics[width=1.9\columnwidth,keepaspectratio,frame]
			{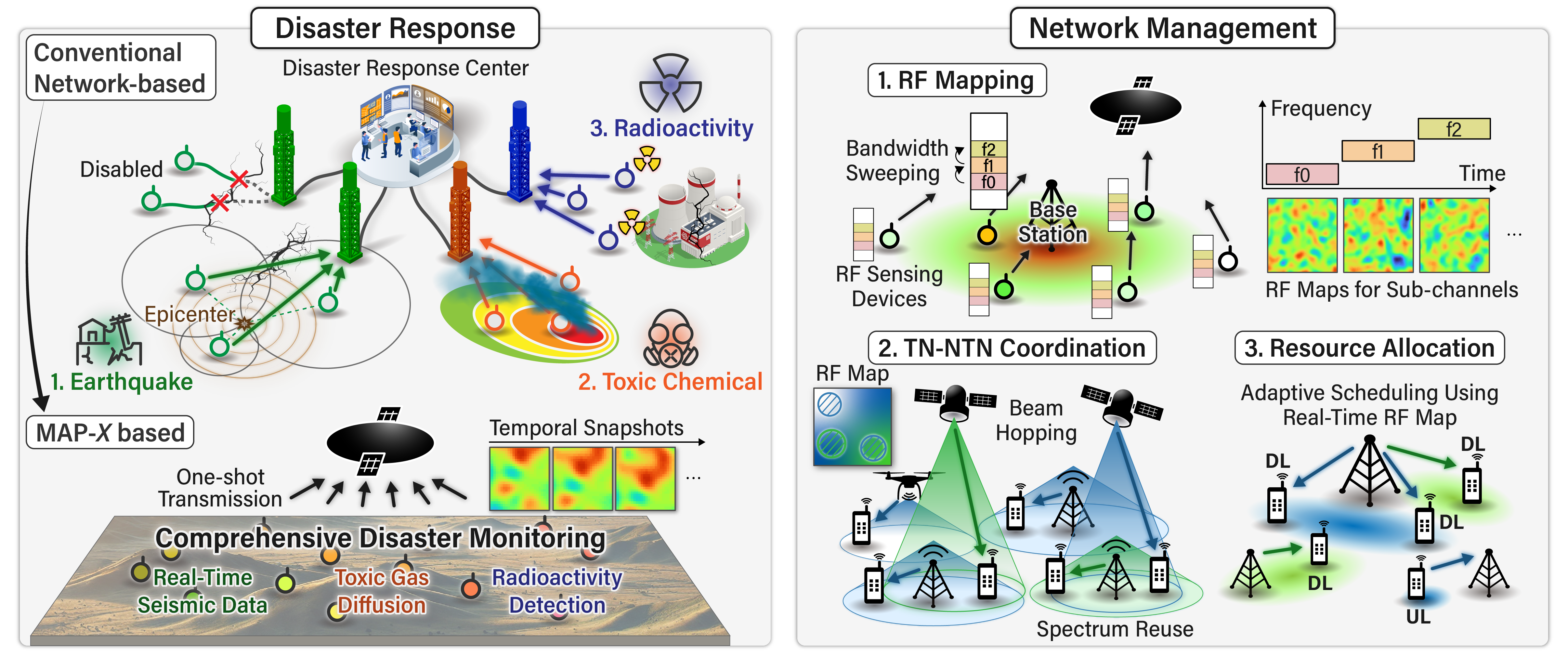}%
			\caption{Potential use cases of MAP-$X$.}
			\label{fig04}
		}
	\end{center}
	\vspace{-15pt}
\end{figure*}

Environmental and industrial monitoring systems aim to measure and collect data related to specific phenomena, requiring wide-area sensing coverage. MAP-$X$, with its extremely short processing times, functions as a highly efficient GIS, capable of collecting data over vast regions and reconstructing high-precision images. This capability makes MAP-$X$ particularly advantageous in scenarios that demand rapid responses based on comprehensive data collection, or when the data exhibit high spatial correlations but low temporal correlations. This section explores potential use cases and future research directions for MAP-$X$, as illustrated in Fig.~\ref{fig04}.

\subsection{Disaster Response}

Disaster scenarios demand immediate and coordinated responses, where latency in data collection can lead to severe impacts. Although various standalone systems have been developed for disaster management, their ability to simultaneously gather hazard signals over large geographic areas is often limited by network constraints. MAP-$X$ can rapidly detect and aggregate signals from vast regions, enabling accurate predictions and real-time assessments across the affected area.

\subsubsection{Earthquake}


For disasters that do not emit electromagnetic waves, such as earthquakes, real-time data collection depends on ground-based sensor networks. However, reliable communication through wired infrastructure is often compromised during such events. Existing monitoring systems primarily detect seismic wave patterns to estimate epicenter, propagation, and magnitude~\cite{earthquakeref}, while detailed propagation analysis is typically performed post-event. MAP-$X$ can enhance this process by aggregating seismic data from a large number of distributed sensors into a wide-area image, enabling faster decision-making and improved forecasting with the support of AI and extensive post-event databases.

\subsubsection{Toxic Chemicals}


Toxic gases can spread rapidly, influenced by chemical and environmental conditions. While certain gases (e.g., sulfur dioxide, ozone) are detectable via specific electromagnetic signatures, many remain invisible to remote sensing platforms~\cite{gasref}. Diffusion models can simulate dispersion but rely on timely environmental input, which is often unavailable in real time. MAP-$X$ enables continuous tracking of concentration changes over both time and space, offering real-time biochemical maps that are essential for identifying safety zones that cannot be reliably predicted using chemical diffusion models or non-real-time sensing data.

\subsubsection{Radioactivity}


Since radiation cannot be detected through traditional remote sensing, ground-based sensors remain essential for monitoring radioactive events. Spatial estimation techniques like kriging can reconstruct radiation levels, but rapid, wide-area coverage is difficult to achieve~\cite{krigingradioref}. Systems like RadNet~\cite{radnetref} offer real-time detection but rely heavily on dense infrastructure. MAP-$X$ provides a low-latency alternative by directly collecting data from a large number of distributed sensors, enabling fast and accurate radiation mapping critical to minimizing exposure risk.

	\subsection{Network Management}
	
A key trend in the evolution toward 6G is the ability to quickly collect and process extensive datasets for AI~\cite{aitrendref}. However, due to the cell-based architecture, current wireless networks are limited in gathering and utilizing spatiotemporal radio data over large geographical areas. MAP-$X$ can support rapid data collection that directly enhances wireless resource management by providing critical information to base stations with extremely low latency.

	\subsubsection{RF Mapping}

Traditional RF mapping, generated from data collected at specific locations, reflect the long-term characteristics of RF signals transmitted by base stations or Wi-Fi routers~\cite{rfmapref}. These maps require significant time to develop and are primarily useful for medium- to long-term network planning, such as base station deployment or policy optimization. However, they are not suited for real-time service optimization through dynamic coordination between base stations. By enabling wide-area data reconstruction at the subframe level, MAP-$X$ offers the ability to optimize resource utilization at the radio resource control (RRC) layer and within physical channel scheduling.

	\subsubsection{Interference Management for NTN}

Recent advancements in satellite miniaturization and reduced launch costs have paved the way for a new era of mobile broadband services enabled by non-terrestrial platforms. This shift has led to integrated models where terrestrial networks (TN) coexist with NTN~\cite{myjsac}. However, differences in the operational characteristics between these networks complicate seamless coordination. Consequently, extensive research has focused on spectrum allocation and interoperability between the two systems~\cite{iftemperatureref}. MAP-$X$’s ability to collect real-time, wide-area data can bring TN-NTN coordination closer to theoretical performance limits. By generating real-time RF maps over extensive areas, MAP-$X$ enables the identification of unused spectrum in specific terrestrial regions, allowing satellites to employ efficient resource allocation and beam hopping.

	\subsubsection{Coordinated Resource Allocation}
	
In conventional cellular communication systems, inter-cell interference arises from uncoordinated scheduling decisions across different cells. To advance towards more sophisticated scheduling, base stations must not only minimize interference within their own cells but also collaboratively optimize resource allocation strategies across multiple cells. The capability of MAP-$X$ to rapidly generate RF maps enables the detection of interference-free opportunities for edge users at the subframe level, allowing for real-time integration into scheduling decisions.

\subsection{Future Research Directions and Challenges}

Future work should focus on optimizing MAP-$X$ under practical system constraints. This includes refining performance under realistic wireless channels, HAPS mobility, and device limitations such as imperfect synchronization and hardware nonidealities. Furthermore, validation through hardware-in-the-loop testing or prototyping is needed to assess feasibility in real-world deployments.

Another key direction is enhancing the AI pipeline. In addition to the AI-based post-processing introduced in this article, AI-based pre-processing can also be applied at the data level to compress multi-dimensional sensor readings into a lower-dimensional latent representation before transmission. This approach is particularly effective in MAP-$X$, where each device transmits only one scalar value per process. Optimizing the cooperation between ground stations and HAPS for efficient data relay and model adaptation will be essential to scaling MAP-$X$ across various use cases.

	\section{Conclusion}
	
In this article, we introduced the integration of AI with Massive Aerial Processing for $X$ (MAP-$X$), an innovative wide-area data reconstruction framework. We compared two deep learning approaches, DNN-based pointwise estimation and CNN-based image reconstruction methods, built on top of the mathematically-driven linear model. While the pointwise estimation method enables fast adaptation to environmental changes through online training, the image reconstruction method achieves higher data reconstruction accuracy by leveraging long-term data collection at terrestrial stations. These AI-enhanced MAP-$X$ systems have the potential to revolutionize data collection and decision-making in latency-critical IoT applications, enabling faster and more accurate responses across various real-world scenarios.

	\section{Acknowledgments}
	
	This work was supported by Institute of Information \& communications Technology Planning \& Evaluation (IITP) grant funded by the MSIT (2022-0-00704 and RS-2024-00428780).

\bibliographystyle{IEEEtran}
\bibliography{Moon_mag_mapx} 	

\begin{thebibliography}{10}
\providecommand{\url}[1]{#1}
\csname url@samestyle\endcsname
\providecommand{\newblock}{\relax}
\providecommand{\bibinfo}[2]{#2}
\providecommand{\BIBentrySTDinterwordspacing}{\spaceskip=0pt\relax}
\providecommand{\BIBentryALTinterwordstretchfactor}{4}
\providecommand{\BIBentryALTinterwordspacing}{\spaceskip=\fontdimen2\font plus
\BIBentryALTinterwordstretchfactor\fontdimen3\font minus
  \fontdimen4\font\relax}
\providecommand{\BIBforeignlanguage}[2]{{%
\expandafter\ifx\csname l@#1\endcsname\relax
\typeout{** WARNING: IEEEtran.bst: No hyphenation pattern has been}%
\typeout{** loaded for the language `#1'. Using the pattern for}%
\typeout{** the default language instead.}%
\else
\language=\csname l@#1\endcsname
\fi
#2}}
\providecommand{\BIBdecl}{\relax}
\BIBdecl

\bibitem{oacref}
G.~Zhu \emph{et~al.}, ``Over-the-air computing for wireless data aggregation in
  massive {IoT},'' \emph{{IEEE} Wireless Commun.}, vol.~28, no.~4, pp. 57--65,
  Sep. 2021.

\bibitem{gisref}
Q.~Meng \emph{et~al.}, ``Assessment of regression kriging for spatial
  interpolation – comparisons of seven {GIS} interpolation methods,''
  \emph{Cartogr. Geographic Inf. Sci.}, vol.~40, no.~1, pp. 28--39, Jan. 2013.

\bibitem{remsenref}
M.~Fiore, ``Full network sensing: Architecting {6G} beyond communications,''
  \emph{IEEE Netw.}, vol.~37, no.~3, pp. 232--239, May/Jun. 2023.

\bibitem{myjref}
\BIBentryALTinterwordspacing
H.-J. Moon \emph{et~al.}, ``{MAP-$X$}: Massive field data processing for
  real-time wide-area mapping using high-altitude platforms with {MIMO},''
  \emph{IEEE Trans. Wireless Commun.}, 2025. [Online]. Available:
  \url{https://www.dropbox.com/scl/fo/0llzu3i2v6lxx7huauf1y/AJWNk5Toq7SMlytI0vgxYXs?rlkey=hyeaj9ix2rylcjj9clzabtuk5&st=f8mpvohd&dl=0}
\BIBentrySTDinterwordspacing

\bibitem{risref}
A.~Azizi and A.~Farhang, ``{RIS} meets aerodynamic {HAPS}: A multi-objective
  optimization approach,'' \emph{{IEEE} Wireless Commun. Lett.}, vol.~12,
  no.~11, pp. 1851--1855, Nov. 2023.

\bibitem{sblueref2}
P.~Zhang \emph{et~al.}, ``Spatial field reconstruction and sensor selection in
  heterogeneous sensor networks with stochastic energy harvesting,''
  \emph{{IEEE} Trans. Signal Process.}, vol.~66, no.~9, pp. 2245--2257, May
  2018.

\bibitem{unetref}
\BIBentryALTinterwordspacing
O.~Ronneberger \emph{et~al.}, ``{U-Net}: Convolutional networks for biomedical
  image segmentation,'' May 2015. [Online]. Available:
  \url{https://arxiv.org/abs/1505.04597}
\BIBentrySTDinterwordspacing

\bibitem{earthquakeref}
G.~Liu \emph{et~al.}, ``Volcanic earthquake timing using wireless sensor
  networks,'' in \emph{Proc. ISPN}, 2013.

\bibitem{gasref}
M.~Faruolo \emph{et~al.}, ``A daytime multisensor satellite system for global
  gas flaring monitoring,'' \emph{IEEE Trans. Geosci. and Remote Sens.},
  vol.~60, pp. 1--17, Jan. 2022.

\bibitem{krigingradioref}
L.~Yi \emph{et~al.}, ``Localized confident information coverage hole detection
  in internet of things for radioactive pollution monitoring,'' \emph{IEEE
  Access}, vol.~5, pp. 18\,665--18\,674, Sep. 2017.

\bibitem{radnetref}
R.~Fraass, ``{RadNet} national air monitoring program,'' in \emph{Nuclear
  Terrorism and National Preparedness}.\hskip 1em plus 0.5em minus 0.4em\relax
  Springer Netherlands, 2015.

\bibitem{aitrendref}
K.~B. Letaief \emph{et~al.}, ``The roadmap to {6G}: {AI} empowered wireless
  networks,'' \emph{{IEEE} Commun. Mag.}, vol.~57, no.~8, pp. 84--90, Aug.
  2019.

\bibitem{rfmapref}
Y.~Li \emph{et~al.}, ``{Geo2SigMap}: High-fidelity {RF} signal mapping using
  geographic databases,'' in \emph{Proc. IEEE DySPAN}, May 2024, pp. 277--285.

\bibitem{myjsac}
H.-J. Moon and C.-B. Chae, ``Cooperative ground-satellite scheduling and power
  allocation for urban air mobility networks,'' \emph{{IEEE} J. Sel. Areas
  Commun.}, vol.~43, no.~1, pp. 218--233, Jan. 2025.

\bibitem{iftemperatureref}
O.~Y. Kolawole \emph{et~al.}, ``On the performance of cognitive
  satellite-terrestrial networks,'' \emph{IEEE Trans. Cogn. Commun. and Netw.},
  vol.~3, no.~4, pp. 668--683, Oct. 2017.

\end{thebibliography}

\end{document}